\documentclass[aps,prb,twocolumn,10pt,preprint,onecolumn]{revtex4-1} 

\usepackage{amssymb,amsmath}
\usepackage{graphicx}

\usepackage{natbib}
\usepackage{multirow}
\usepackage{bm}

\usepackage[utf8]{inputenc}
\usepackage[T1]{fontenc}
\usepackage{mathptmx}

\begin{document}
	\title{Lagrangian based heat conduction}

	\author{Ferenc M\'arkus}
	\email[]{markus@phy.bme.hu}
	\affiliation{Department of Physics,
		Budapest University of Technology and Economics,
		H-1111 Budafoki \'ut 8., Budapest, Hungary}

	\author{Andr\'as Szegleti}
	\email[Corresponding author: ]{andras.szegleti@edu.bme.hu}
	\affiliation{Department of Physics,
		Budapest University of Technology and Economics,
		H-1111 Budafoki \'ut 8., Budapest, Hungary}

	\date{\today}

\begin{abstract}
	Based on the Lagrangian description of the dissipative oscillator, the Hamiltonian description of Fourier heat conduction is treated here. The method enables us to calculate the solution of thermal propagation involving the Maxwell-Cattaneo-Vernotte (MCV) telegrapher type also, in which both the initial and the boundary conditions are taken into account. The presented study offers a new kind of solution to certain partial differential equations.
\end{abstract}

\maketitle

\section{Introduction}
In the past, there were several attempts to deduce dissipative processes within the framework of least action principle (Hamilton's principle)\cite{rayleigh1877theory,PhysRev.37.405, PhysRev.38.2265, RevModPhys.17.343, de1962non, sieniutycz2012conservation}. These descriptions introduce a so-called dissipation function to the Euler--Lagrange equations. However, this additional term cannot be derived from the Lagrangian. Bateman suggested the construction of Lagrangian by doubling the degrees of freedom \cite{PhysRev.38.815}, by which a mirror system is defined. It means such a description in which the energy of the examined system decreases, while, reversely, the energy of the related mirror system increases. \\
A present suggestion for the Lagrangian formulation of dissipative harmonic oscillator, including the initial conditions, is successfully elaborated \cite{szegleti2020dissipation} by the application potential procedure \cite{markus1991variational, gambar1991variational, nyiri1991construction, PhysRevE.50.1227, gambar2016application}. It is shown that the unphysical solutions can be eliminated by a suitable choice of the initial conditions. In this work, this method is applied for heat transfer processes. We assume that the thermal process is described by Maxwell-Cattaneo-Vernotte (MCV) telegrapher equation for the temperature. A damped oscillator-like equation can be obtained by a half-Fourier transform, after which the potential method may be used. The initial conditions come from the Fourier transform of the initial temperature distribution. The temperature field can be constructed using the defining equation of the potential and an inverse Fourier transform. The crucial fact is that the solution is regular, i.e., the divergent solutions are not present. \\
It seems that this procedure gives a new approach to the solutions of linear partial differential equations. Here, the initial conditions and the Dirichlet boundary condition are also involved. It may be interesting simply from mathematical viewpoints.

\section{Damped harmonic oscillator}
The equation of motion for the damped harmonic oscillator is
\begin{equation}
	m\ddot x + c \dot x + k x = 0,
\end{equation}
or in a more suitable writing
\begin{equation}
	m\ddot x + 2m\lambda \dot x + m\omega^2 x = 0,
\end{equation}
where \(m\) is the mass, \(k\) is the spring constant, \(\lambda = c/2m\) is the damping coefficient and \(\omega = \sqrt{k/m}\) is the angular frequency.

The measurable quantity, \(x\), is used to define the potential, \(q\). As the coefficients are constants, the adjoint equation is
\begin{equation}
	x = \ddot q - 2\lambda \dot q + \omega^2 q.
	\label{eq:dlho:potencial_definition}
\end{equation}
By following the method described in Refs.~\cite{szegleti2020dissipation,PhysRevE.50.1227}, the following Lagrangian holds
\begin{equation}
	L = \frac{1}{2} \left( \ddot q - 2\lambda \dot q + \omega^2 q \right)^2.
\end{equation}

\section{Lagrangian of telegrapher's heat conduction}
The equation of motion for the telegrapher's heat transport (a la Maxwell-Cattaneo-Vernotte) is ($T(x,t)$)
\begin{equation}
	\tau \ddot T + \varrho c_v \dot T - \lambda' \Delta T = 0,
\end{equation}
taking the half-Fourier transform
\begin{equation}
	\Delta T \longrightarrow k^2 \, \tilde{T}(k,t)
\end{equation}
thus

\begin{equation}
	\ddot{\tilde{T}} + \frac{\varrho c_v}{\tau} \dot{\tilde{T}} + \frac{\lambda'}{\tau} k^2 \, \tilde{T} = 0
\end{equation}
introducing
\begin{equation}
	2 \lambda = \frac{\varrho c_v}{\tau}
\end{equation}

\begin{equation}
	\omega^2 = \frac{\lambda'}{\tau} k^2
\end{equation}

\begin{equation}
	\ddot{\tilde{T}} + 2 \lambda \dot{\tilde{T}} + \omega^2 \, \tilde{T} = 0
\end{equation}

\begin{equation}
	\tilde{T} = \ddot{\tilde{q}} - 2\lambda \dot{\tilde{q}} + \omega^2 \tilde{q}  \label{half-Fourier_temperature_definition}
\end{equation}

$\tilde{q}(k,t)$ is half-Fourier transformed of $q(x,t)$; Fourier heat conduction is an overdamped case $\lambda > \omega >> 1$

Since we have a similar formulation to the damped harmonic oscillator, thus the following Lagrangian can be written:

\begin{equation}
	L = \frac{1}{2} \left( \ddot{\tilde{q}} - 2\lambda \dot{\tilde{q}} + \omega^2 \tilde{q} \right)^2.
\end{equation}

\section{Solution in Fourier's space}
For the sake of simlicity, we focus just on the solutions that pertain to the overdamped cases. As Szegleti et al. \cite{szegleti2020dissipation} showed the solution for \(\tilde{q}(t)\) can be easily calculated as
\begin{eqnarray}
	\tilde{q}(k,t) &=&
	a_1\mathrm{e}^{-(\lambda+\gamma)t}
	+ a_2\mathrm{e}^{-(\lambda-\gamma)t} \nonumber \\
	& &+ b_1\mathrm{e}^{(\lambda+\gamma)t}
	+ b_2\mathrm{e}^{(\lambda-\gamma)t},
	\label{thermal_potential_solution}
\end{eqnarray}
where \(\gamma = \sqrt{\lambda^2-\omega^2}\). The terms proportional to \(\mathrm{e}^{\lambda t}\) are solutions of the adjoint operator, hence they have non-physical meaning, consequently, they could not have role in formulation of the measurable \(\tilde{q}(k,t)\).

The initial conditions for the measurable quantities can be expressed as usual
\begin{eqnarray}
	\tilde{T}(k,t=0) &=& T_0(k), \\[4pt]
	\dot{\tilde{T}}(k,t=0) &=& S_0(k).
\end{eqnarray}

Applying a suitable choice of the initial conditions for the potential
\begin{eqnarray}
	{\tilde{q}}(k,t=0) &=& \frac{2\lambda T_0(k) + S_0(k)}{4\lambda(\lambda^2-\gamma^2)}, \\[4pt]
	\dot{\tilde{q}}(k,t=0) &=& -\frac{T_0(k)}{4\lambda}, \\[4pt]
	\ddot{\tilde{q}}(k,t=0) &=& -\frac{S_0(k)}{4\lambda}, \\[4pt]
	\dddot{\tilde{q}}(k,t=0) &=& \frac{(\lambda^2-\gamma^2T_0(k) + 2\lambda S_0(k)}{4\lambda},
\end{eqnarray}
it can be clearly proved that the non-physical solutions (the exponentially increasing terms in Eq.~(\ref{thermal_potential_solution})) will vanish, thus~the coefficients will be
\begin{eqnarray}
	a_1 &=& \frac{(\gamma-\lambda)T_0(k) - S_0(k)}{8\gamma\lambda(\lambda+\gamma)},\\[4pt]
	a_2 &=& \frac{(\gamma+\lambda)T_0(k) + S_0(k)}{8\gamma\lambda(\lambda-\gamma)},\\[4pt]
	b_1 &=& 0,\\[4pt]
	b_2 &=& 0.
\end{eqnarray}

Finally, the potential -- fitted to the initial conditions -- can be obtained in the form
\begin{eqnarray}
	\tilde{q}(k,t) &=&
	\frac{(\gamma-\lambda)T_0(k) - S_0(k)}{8\gamma\lambda(\lambda+\gamma)} \mathrm{e}^{-(\lambda+\gamma)t} \nonumber \\
	& &+\frac{(\gamma+\lambda)T_0(k) + S_0(k)}{8\gamma\lambda(\lambda-\gamma)} \mathrm{e}^{-(\lambda-\gamma)t}.
	\label{potential_solution_final}
\end{eqnarray}

The half-Fourier temperature $\tilde{T}(k,t)$ can be calculated by Eq.~(\ref{half-Fourier_temperature_definition}).

\section{Results}

\begin{figure}[h]
	\centering
	\input{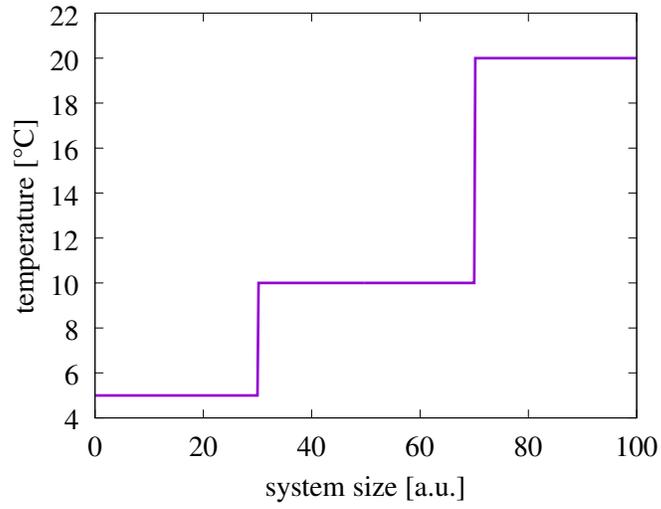}
	\caption{Initial temperature distribution.}
	\label{fig:Fourier_initial_temperature_distribution}
\end{figure}

A one-dimensional silicon bar is put in the range $30 \leq x \leq 70$. The initial temperature distribution is shown in Fig.~\ref{fig:Fourier_initial_temperature_distribution}. The left hand side boundary is in the range $0 \leq x \leq 30$ with the fixed temperature $T_{\textrm{left}}(30) = 5$ $^0$C, the right hand side boundary is in the range $70 \leq x \leq 100$ with the fixed temperature $T_{\textrm{right}}(70) = 20$ $^0$C. The temperature at the boundaries can be fixed by manipulating certain material parameters, for example, by setting a huge mass density. In this case, the change of the temperature would require an enormous internal energy transfer. The described situation fits the Dirichlet boundary condition. As is discussed above, the Fourier transform of the initial condition is needed, which can be formulated as
\begin{eqnarray}
	T_0(k) &=&
	\frac{5(-\textrm{i} + \textrm{i} e^{-0.3 \textrm{i} k})}{k}  \nonumber \\
	& &- \frac{10(\textrm{i}e^{-0.3 \textrm{i} k} - \textrm{i} e^{-0.7 \textrm{i} k})}{k} \nonumber \\
	& &+ \frac{20(-\textrm{i}e^{-0.7 \textrm{i} k} + \textrm{i} e^{- \textrm{i} k})}{k} .
\end{eqnarray}

$S_0(k)$ would relate to the initial velocity of the oscillator, but there is no identifiable role of it, thus
$S_0(k) = 0$ is taken. The calculations go numerically. \\

The applied parameters for silicon: mass density $\varrho = 2300$ kg/m$^3$, heat conductivity $\lambda = 149$ W/m $\cdot$ K and spcific heat $c_v =$ 700 J/kg\,K. The relaxation parameter $\tau$ is taken enough small to achive the diffusive solution: $\tau = 10^{-4}$ 1/s. \\

The obtained graphs at different times are the following:%
\begin{figure}[h]
	\centering
	\input{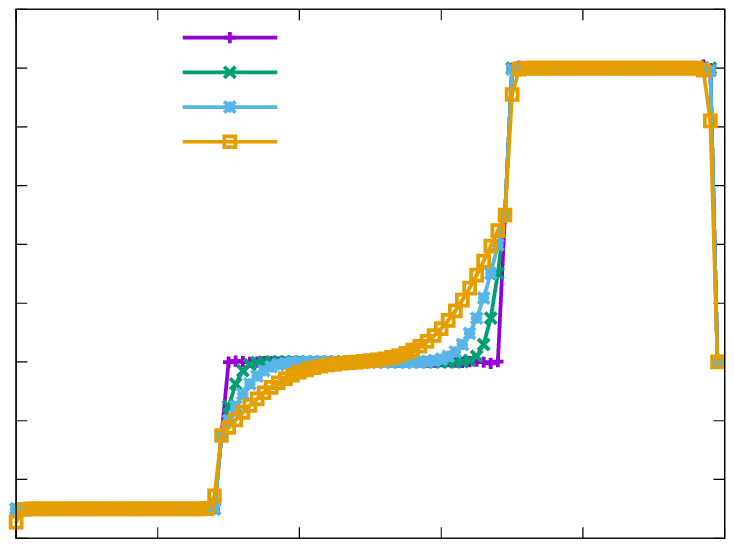}
	\caption{Temperature distribution at \( t=0\,\mathrm{s}, ~2\,\mathrm{s}, ~8\,\mathrm{s}, ~30\,\mathrm{s} \).}
	\label{fig:temperature_distribution_t=0-30}
\end{figure}

\section{Conclusion}
We pointed out that elaborated Lagrangian formulation of the dissipative oscillator can be successfully applied for such irreversible process as thermal propagation, however, not only for the Fourier's heat conduction but also for the Maxwell-Cataneo-Vernotte case. The generalization procedure is based on the similarity between the equations of the dissipative oscillator and the half-Fourier MCV. The Lagrange-Hamilton framework can be widened for dissipative systems, obtaining convergent solutions. Furthermore, we would like to draw attention to this potential-based solution of linear partial differential equations. The method works independently the fact that the differential equation pertains to either the non-dissipative or dissipative process.

\section{Acknowledgments}
We acknowledge the support of the NKFIH Grant nos. K119442 and 2017-1.2.1-NKP-2017-00001. The research has also been supported by the NKFIH Fund (TKP2020 IES, grant no. BME-IE-NAT) based on the charter of bolster issued by the NKFIH Office under the auspices of the Ministry for Innovation and Technology.


\bibliography{LagrangianHeatConduction_refs}

\begin{thebibliography}{13}%
\makeatletter
\providecommand \@ifxundefined [1]{%
 \@ifx{#1\undefined}
}%
\providecommand \@ifnum [1]{%
 \ifnum #1\expandafter \@firstoftwo
 \else \expandafter \@secondoftwo
 \fi
}%
\providecommand \@ifx [1]{%
 \ifx #1\expandafter \@firstoftwo
 \else \expandafter \@secondoftwo
 \fi
}%
\providecommand \natexlab [1]{#1}%
\providecommand \enquote  [1]{``#1''}%
\providecommand \bibnamefont  [1]{#1}%
\providecommand \bibfnamefont [1]{#1}%
\providecommand \citenamefont [1]{#1}%
\providecommand \href@noop [0]{\@secondoftwo}%
\providecommand \href [0]{\begingroup \@sanitize@url \@href}%
\providecommand \@href[1]{\@@startlink{#1}\@@href}%
\providecommand \@@href[1]{\endgroup#1\@@endlink}%
\providecommand \@sanitize@url [0]{\catcode `\\12\catcode `\$12\catcode
  `\&12\catcode `\#12\catcode `\^12\catcode `\_12\catcode `\%12\relax}%
\providecommand \@@startlink[1]{}%
\providecommand \@@endlink[0]{}%
\providecommand \url  [0]{\begingroup\@sanitize@url \@url }%
\providecommand \@url [1]{\endgroup\@href {#1}{\urlprefix }}%
\providecommand \urlprefix  [0]{URL }%
\providecommand \Eprint [0]{\href }%
\providecommand \doibase [0]{http://dx.doi.org/}%
\providecommand \selectlanguage [0]{\@gobble}%
\providecommand \bibinfo  [0]{\@secondoftwo}%
\providecommand \bibfield  [0]{\@secondoftwo}%
\providecommand \translation [1]{[#1]}%
\providecommand \BibitemOpen [0]{}%
\providecommand \bibitemStop [0]{}%
\providecommand \bibitemNoStop [0]{.\EOS\space}%
\providecommand \EOS [0]{\spacefactor3000\relax}%
\providecommand \BibitemShut  [1]{\csname bibitem#1\endcsname}%
\let\auto@bib@innerbib\@empty
\bibitem [{\citenamefont {Rayleigh}(1877)}]{rayleigh1877theory}%
  \BibitemOpen
  \bibfield  {author} {\bibinfo {author} {\bibfnamefont {J.}~\bibnamefont
  {Rayleigh}},\ }\href {https://books.google.hu/books?id=GyI5AAAAMAAJ} {\emph
  {\bibinfo {title} {The Theory of Sound}}},\ \bibinfo {series} {The Theory of
  Sound}\ No.\ \bibinfo {number} {1. k.}\ (\bibinfo  {publisher} {Macmillan and
  Company},\ \bibinfo {year} {1877})\BibitemShut {NoStop}%
\bibitem [{\citenamefont {Onsager}(1931{\natexlab{a}})}]{PhysRev.37.405}%
  \BibitemOpen
  \bibfield  {author} {\bibinfo {author} {\bibfnamefont {L.}~\bibnamefont
  {Onsager}},\ }\href {\doibase 10.1103/PhysRev.37.405} {\bibfield  {journal}
  {\bibinfo  {journal} {Phys. Rev.}\ }\textbf {\bibinfo {volume} {37}},\
  \bibinfo {pages} {405} (\bibinfo {year} {1931}{\natexlab{a}})}\BibitemShut
  {NoStop}%
\bibitem [{\citenamefont {Onsager}(1931{\natexlab{b}})}]{PhysRev.38.2265}%
  \BibitemOpen
  \bibfield  {author} {\bibinfo {author} {\bibfnamefont {L.}~\bibnamefont
  {Onsager}},\ }\href {\doibase 10.1103/PhysRev.38.2265} {\bibfield  {journal}
  {\bibinfo  {journal} {Phys. Rev.}\ }\textbf {\bibinfo {volume} {38}},\
  \bibinfo {pages} {2265} (\bibinfo {year} {1931}{\natexlab{b}})}\BibitemShut
  {NoStop}%
\bibitem [{\citenamefont {Casimir}(1945)}]{RevModPhys.17.343}%
  \BibitemOpen
  \bibfield  {author} {\bibinfo {author} {\bibfnamefont {H.~B.~G.}\
  \bibnamefont {Casimir}},\ }\href {\doibase 10.1103/RevModPhys.17.343}
  {\bibfield  {journal} {\bibinfo  {journal} {Rev. Mod. Phys.}\ }\textbf
  {\bibinfo {volume} {17}},\ \bibinfo {pages} {343} (\bibinfo {year}
  {1945})}\BibitemShut {NoStop}%
\bibitem [{\citenamefont {de~Groot}\ \emph {et~al.}(1962)\citenamefont
  {de~Groot}, \citenamefont {Groot},\ and\ \citenamefont {Mazur}}]{de1962non}%
  \BibitemOpen
  \bibfield  {author} {\bibinfo {author} {\bibfnamefont {S.}~\bibnamefont
  {de~Groot}}, \bibinfo {author} {\bibfnamefont {S.}~\bibnamefont {Groot}}, \
  and\ \bibinfo {author} {\bibfnamefont {P.}~\bibnamefont {Mazur}},\ }\href
  {https://books.google.hu/books?id=3b-wAAAAIAAJ} {\emph {\bibinfo {title}
  {Non-equilibrium Thermodynamics}}},\ Dover books on physics and chemistry\
  (\bibinfo  {publisher} {North-Holland Publishing Company},\ \bibinfo {year}
  {1962})\BibitemShut {NoStop}%
\bibitem [{\citenamefont {Sieniutycz}(2012)}]{sieniutycz2012conservation}%
  \BibitemOpen
  \bibfield  {author} {\bibinfo {author} {\bibfnamefont {S.}~\bibnamefont
  {Sieniutycz}},\ }\href@noop {} {\emph {\bibinfo {title} {Conservation laws in
  variational thermo-hydrodynamics}}},\ Vol.\ \bibinfo {volume} {279}\
  (\bibinfo  {publisher} {Springer Science \& Business Media},\ \bibinfo {year}
  {2012})\BibitemShut {NoStop}%
\bibitem [{\citenamefont {Bateman}(1931)}]{PhysRev.38.815}%
  \BibitemOpen
  \bibfield  {author} {\bibinfo {author} {\bibfnamefont {H.}~\bibnamefont
  {Bateman}},\ }\href {\doibase 10.1103/PhysRev.38.815} {\bibfield  {journal}
  {\bibinfo  {journal} {Phys. Rev.}\ }\textbf {\bibinfo {volume} {38}},\
  \bibinfo {pages} {815} (\bibinfo {year} {1931})}\BibitemShut {NoStop}%
\bibitem [{\citenamefont {Szegleti}\ and\ \citenamefont
  {M{\'a}rkus}(2020)}]{szegleti2020dissipation}%
  \BibitemOpen
  \bibfield  {author} {\bibinfo {author} {\bibfnamefont {A.}~\bibnamefont
  {Szegleti}}\ and\ \bibinfo {author} {\bibfnamefont {F.}~\bibnamefont
  {M{\'a}rkus}},\ }\href@noop {} {\bibfield  {journal} {\bibinfo  {journal}
  {Entropy}\ }\textbf {\bibinfo {volume} {22}},\ \bibinfo {pages} {930}
  (\bibinfo {year} {2020})}\BibitemShut {NoStop}%
\bibitem [{\citenamefont {M{\'a}rkus}\ and\ \citenamefont
  {Gamb{\'a}r}(1991)}]{markus1991variational}%
  \BibitemOpen
  \bibfield  {author} {\bibinfo {author} {\bibfnamefont {F.}~\bibnamefont
  {M{\'a}rkus}}\ and\ \bibinfo {author} {\bibfnamefont {K.}~\bibnamefont
  {Gamb{\'a}r}},\ }\href@noop {} {\bibfield  {journal} {\bibinfo  {journal}
  {Journal of Non-Equilibrium Thermodynamics}\ }\textbf {\bibinfo {volume}
  {16}},\ \bibinfo {pages} {27} (\bibinfo {year} {1991})}\BibitemShut {NoStop}%
\bibitem [{\citenamefont {Gamb{\'a}r}\ \emph {et~al.}(1991)\citenamefont
  {Gamb{\'a}r}, \citenamefont {M{\'a}rkus},\ and\ \citenamefont
  {Ny\'{\i}ri}}]{gambar1991variational}%
  \BibitemOpen
  \bibfield  {author} {\bibinfo {author} {\bibfnamefont {K.}~\bibnamefont
  {Gamb{\'a}r}}, \bibinfo {author} {\bibfnamefont {F.}~\bibnamefont
  {M{\'a}rkus}}, \ and\ \bibinfo {author} {\bibfnamefont {B.}~\bibnamefont
  {Ny\'{\i}ri}},\ }\href@noop {} {\bibfield  {journal} {\bibinfo  {journal}
  {Journal of Non-Equilibrium Thermodynamics}\ }\textbf {\bibinfo {volume}
  {16}},\ \bibinfo {pages} {217} (\bibinfo {year} {1991})}\BibitemShut
  {NoStop}%
\bibitem [{\citenamefont {Ny\'{\i}ri}(1991)}]{nyiri1991construction}%
  \BibitemOpen
  \bibfield  {author} {\bibinfo {author} {\bibfnamefont {B.}~\bibnamefont
  {Ny\'{\i}ri}},\ }\href@noop {} {\bibfield  {journal} {\bibinfo  {journal}
  {Journal of Non-Equilibrium Thermodynamics}\ }\textbf {\bibinfo {volume}
  {16}},\ \bibinfo {pages} {39} (\bibinfo {year} {1991})}\BibitemShut {NoStop}%
\bibitem [{\citenamefont {Gamb\'ar}\ and\ \citenamefont
  {M\'arkus}(1994)}]{PhysRevE.50.1227}%
  \BibitemOpen
  \bibfield  {author} {\bibinfo {author} {\bibfnamefont {K.}~\bibnamefont
  {Gamb\'ar}}\ and\ \bibinfo {author} {\bibfnamefont {F.}~\bibnamefont
  {M\'arkus}},\ }\href {\doibase 10.1103/PhysRevE.50.1227} {\bibfield
  {journal} {\bibinfo  {journal} {Phys. Rev. E}\ }\textbf {\bibinfo {volume}
  {50}},\ \bibinfo {pages} {1227} (\bibinfo {year} {1994})}\BibitemShut
  {NoStop}%
\bibitem [{\citenamefont {Gamb{\'a}r}\ \emph {et~al.}(2016)\citenamefont
  {Gamb{\'a}r}, \citenamefont {Lendvay}, \citenamefont {Lovassy},\ and\
  \citenamefont {Bugyj{\'a}s}}]{gambar2016application}%
  \BibitemOpen
  \bibfield  {author} {\bibinfo {author} {\bibfnamefont {K.}~\bibnamefont
  {Gamb{\'a}r}}, \bibinfo {author} {\bibfnamefont {M.}~\bibnamefont {Lendvay}},
  \bibinfo {author} {\bibfnamefont {R.}~\bibnamefont {Lovassy}}, \ and\
  \bibinfo {author} {\bibfnamefont {J.}~\bibnamefont {Bugyj{\'a}s}},\
  }\href@noop {} {\bibfield  {journal} {\bibinfo  {journal} {Acta Polytechnica
  Hungarica}\ }\textbf {\bibinfo {volume} {13}},\ \bibinfo {pages} {173}
  (\bibinfo {year} {2016})}\BibitemShut {NoStop}%
\end{thebibliography}%

\end{document}